\documentclass[aps,prl,reprint,superscriptaddress]{revtex4-2}
\usepackage[dvipsnames]{xcolor}
\usepackage{amsmath,bm}
\usepackage{wrapfig}
\usepackage{graphicx}

\newcommand{\qq}{{\bm{q}}}
\newcommand{\pp}{{\bm{p}}}
\newcommand{\RR}{{\bm{R}}}
\newcommand{\ii}{{\bm{r}}}
\newcommand{\jj}{{\bm{j}}}
\newcommand{\kk}{{\bm{k}}}

\begin{document}

\title{Thermal conductivity in noncollinear magnets}
\author{Margherita Parodi}
\affiliation{Dipartimento di Fisica, Università di Genova, Genova 16146, Italy}
\affiliation{Istituto Italiano di Tecnologia, Genova 16163, Italy}
\author{Sergey Artyukhin}
\affiliation{Quantum Materials Theory, Genova 16162, Italy}

\begin{abstract}
    Magnetic memory and logic devices, including prospective ones based on skyrmions, inevitably produce heat. Thus, controlling heat flow is essential for their performance. Here we study how non-collinear spin arrangement affects the magnon contribution to thermal conductivity. As a paradigm system, we consider the most basic non-collinear magnet with a spin spiral ground state. Spin noncollinearity leads to anharmonic terms, resulting in magnon fusion and decay processes. These processes determine the magnon lifetime, which can be used to estimate thermal conductivity in a single-mode approximation. However, by solving the full Boltzmann equation numerically, we find a much higher thermal conductivity. This signifies that heat is carried not by individual magnons but by their linear combinations -- relaxons. The thermal conductivity is found to increase with the diminishing spiral pitch, consistent with recent experiments. The results provide the blueprint for calculating magnetic thermal transport in  non-collinear magnets.
\end{abstract}
\maketitle

\paragraph{Introduction ---}

Competing spin interactions, or magnetic frustration, are an ideal playground for noncollinear magnetic states, such as spin spirals and skyrmions. These textures are actively discussed as a potential platform for information technology and are present in many materials, for example, due to competing nearest- and next-nearest-neighbor exchange interactions in rare earth manganites, spin-orbit-driven anisotropic interactions, or thermal fluctuations in MnSi. Spin spirals break inversion symmetry and may give rise to ferroelectric polarization, which holds promise for electric control of magnetism \cite{Katsura2005,Mostovoy2006,Sergienko2006,cheong2007multiferroics}.

In magnetic logic and memory devices, energy is mostly dissipated as magnons. Thus, controlling magnon thermal transport is crucial for keeping the devices operating. This may be particularly important in the light of recent proposals to implement quantum bits via skyrmions and magnetic domain walls \cite{Psaroudaki21,Psaroudaki23,Loss25}.

Recent ultrafast experiments observed strong changes in thermal conductivity (exceeding 50\%) across the phase transition from ferromagnetic to spiral and skyrmion states \cite{slowdown}. The origin of the effect is still debated, and one possible explanation is the scattering of magnons and phonons off the domain walls formed in abundance in non-collinear states \cite{slowdown,ParodiGenetic}.

Here we explore an intrinsic contribution to the thermal conductivity originating from spin noncollinearity, which results in three-magnon anharmonic terms in the Hamiltonian, describing magnon fusion and decay. We find that the higher the pitch of the spiral spin structure, the lower the thermal conductivity. Recent developments in phonon transport have revealed the inadequacy of solving the kinetic equation beyond the single-mode approximation \cite{Hardy70, Cepellotti16}, leading one to consider relaxons, the eigenvectors of the collision matrix, that describe the dissipation of energy flux. We analyze the thermal conductivity within the relaxon framework and find that most of the thermal conductivity is contributed by only a few relaxons for a wide range of model parameters. The full collision matrix calculation results in orders of magnitude larger thermal conductivity than the single-mode magnon estimate. To the best of our knowledge, this work provides the first implementation of the relaxon theory for the thermal conductivity of a magnetic material.

This Letter is structured as follows: we introduce the magnetic Hamiltonian of a spiral magnet, use Bogoliubov transformation to diagonalize its quadratic part and describe the magnons. The resulting Hamiltonian, with third-order terms, responsible for magnon fusion and decay, is used to formulate the kinetic equation. Its solution is obtained in the basis of eigenvectors of the symmetrized scattering matrix \cite{Hardy70} (the relaxon basis \cite{Cepellotti16}). We compute the thermal conductivity, investigate its dependence on temperature, spiral period, and magnetic anisotropy, and discuss relaxons that account for the principal contributions to the thermal conductivity.

\begin{figure*}[t]
    \centering    \includegraphics[width=\linewidth]{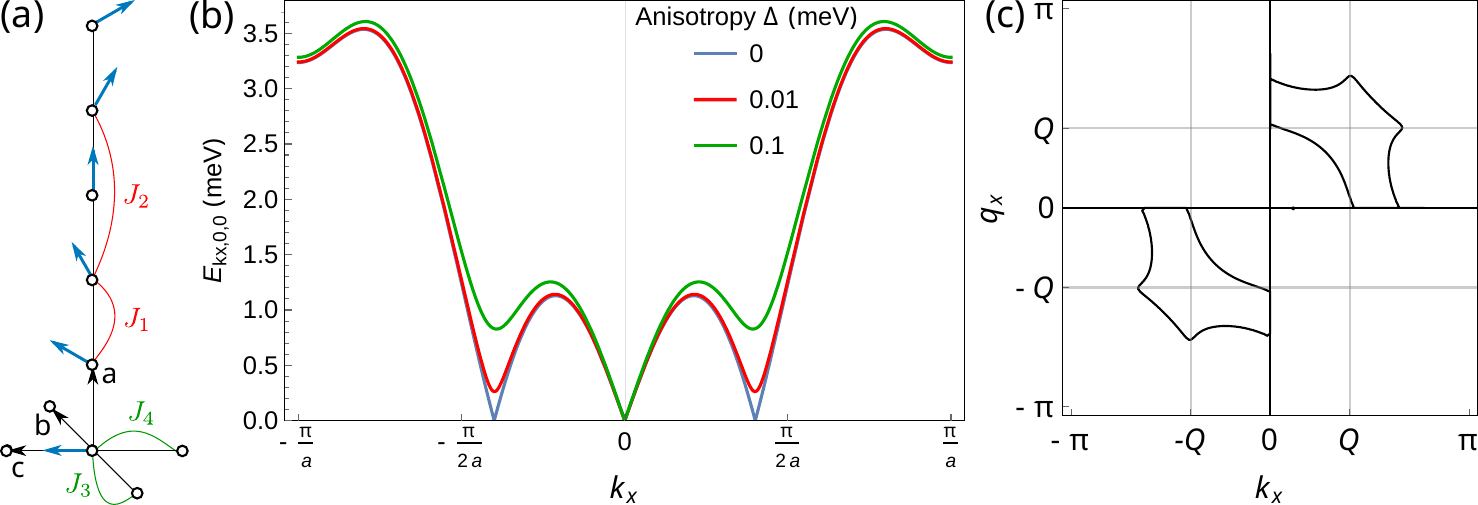}
    \caption{(a) Quasi-1D spin spiral state is stabilized by competing nearest- and next-nearest neighbor exchange interactions $J_1,J_2$ and ferromagnetic interactions $J_3,J_4$. (b) Magnon dispersion along the spiral wave vector $E(k_x,k_y=0,k_z=0)$. The model parameters are $J_2=1$~meV, $Q=\frac{2\pi}{5}$, $J_1=-4J_2\cos Q$ and the strength of the easy plane anisotropy $\Delta$ is indicated in the legend. The easy plane anisotropy favors one plane for the spiral (in our case, the $xz$ plane), so the Goldstone modes associated with the continuous rotational symmetry, broken by the anisotropy, are now gapped. (c) $k_y=k_z=q_y=q_z=0$ cut of the surface in $(\bm{k,q})$ space where the energy conservation law for the scattering event $\epsilon_{\bm{k}}+\epsilon_{\bm{q}}=\epsilon_{\bm{k}+\bm{q}}$ is satisfied.}
    \label{fig:fig1}
\end{figure*}



\paragraph{Model ---}
Non-collinear spin textures emerge naturally in many frustrated systems, and the simplest model that leads to a spin spiral state includes competing interactions between nearest neighbor ($J_1<0$) and next-nearest neighbor ($J_2>0$) spins along the $a$ axis. Such a mechanism with spiral state stabilized by the competition of nearest and next-nearest neighbor exchange interactions is realized in a variety of magnetic materials, including multiferroic CuO, TbMnO$_3$, MnWO$_4$, Ni$_3$V$_2$O$_8$ and monolayer materials NiI$_2$, VI$_2$ and NiBr$_2$, which are considered as a prospective platform for information technology and spintronics \cite{Manipatruni2018}.
We study a quasi-one-dimensional spin spiral state in a Heisenberg model on a cubic lattice, where competing $J_1,J_2$ interactions along the $a$ axis are supplemented with ferromagnetic nearest-neighbor exchange $J_3<0$ in the directions $b$ and $c$ perpendicular to the chain.
We also include an easy plane anisotropy, $\Delta>0$, which favors spins to lie in the $ac$ plane, as seen in Fig.~\ref{fig:fig1}(a). 
Hence, the Hamiltonian is as follows:
\begin{equation}
    \begin{split}
        H=\sum_{\ii} \sum_{\delta=1,2}& \left(  J_\delta\mathbf{S}_{\ii}\cdot \mathbf{S}_{\ii+\mathbf{a}\delta} \right)+\Delta \left(S_{\ii}^y\right)^2\\& +J_3(\mathbf{S}_{\ii}\cdot \mathbf{S}_{\ii+\mathbf{b}}+ \mathbf{S}_{\ii}\cdot \mathbf{S}_{\ii+\mathbf{c}}),
        \label{eq:Ham}
    \end{split}
\end{equation}
where $\mathbf{S}_{\ii}$ is the spin ($S\gg \frac{1}{2}$) on the site $\ii$; $\mathbf{a},\mathbf{b},\mathbf{c}$ are the translation vectors along the Cartesian directions (Fig.~\ref{fig:fig1}(a)). We take the next-nearest neighbor interaction constant along the spiral wave vector $J_2=1$ meV, exchange constants in the perpendicular directions $J_3=-0.1$ meV; the easy-plane anisotropy is $K_z=0.05$ meV. In order to study the $Q$-vector dependence of the thermal conductivity, we use various values of $J_1$, which determine the spiral wave vector via $\cos Q=-J_1/4J_2$.

In a collinear spin structure, such as a ferromagnet or an antiferromagnet, the magnon Hamiltonian has no third order terms. However, non-collinearity of spins leads to cubic anharmonic terms \cite{Dmitriev06}. Since commutation relations of spin operators are neither bosonic nor fermionic, Holstein-Primakoff transformation to bosonic operators $a, a^\dag$, is often used to diagonalize the spin Hamiltonian \cite{HP40}. To the lowest order in magnon occupations, $S_x$ and $S_y$ are expressed via single bosonic operators, $S_x+iS_y\propto a, S_x-iS_y\propto a^\dag$ and $S_z$ – via $a^\dag a$. For collinear spins the quantization axes are parallel and therefore the terms in the exchange operator involving $S_x S_x'$ and $S_y S_y'$ contain two bosonic operators, while $S_zS_z'$ term contains four. In a non-collinear state, however, quantization axes of neighboring spins are tilted with respect to each other, which leads to three-magnon terms, e.g. $\propto \sin Q a_i(S-a_j^\dag a_j)$ (see SI for details). These terms describe events where a magnon splits in two or where two magnons merge into one. They lead to magnon scattering and a finite magnon mean free path, and thus a finite magnetic contribution to thermal conductivity.
Performing the Bogoliubov transformation \cite{Bogolyubov1958,Valatin1958} $a_{\kk} \rightarrow \alpha_{\kk} u_{\kk}-\alpha^\dag_{-\kk} v_{\kk}$, we get to the quadratic part of the Hamiltonian:
\begin{equation}
     H^{(2)}=S\sum_{\kk} \sqrt{A(\kk)^2-B(\kk)^2}\alpha^\dag_{\kk}\alpha_{\kk},\label{eq:H2}
     \end{equation}
where 
\begin{equation}                
    \begin{split}
        A(\kk)=&\sum_{\delta=1,2}\left( J_\delta\left[\left(1+\cos{Q\delta}\right)\cos{k_xa \delta}-2\cos{Q \delta}\right]\right)+\\  
        &+2J_3\left(\cos{k_yb}+\cos{k_zc}-2\right)+\Delta,  \\
        B(\kk) =& \sum_{\delta=1,2}\left( J_\delta\left(\cos{Q \delta}-1\right)\cos{k_xa \delta}\right)-\Delta,
    \end{split}
\end{equation}
and the Bogoliubov coefficient $\theta_\kk$ is defined as
\begin{equation}
      \tanh{2\theta_\kk}=\frac{B(\kk)}{A(\kk)}.
      \label{eq:bogoliubov}
\end{equation}
The quadratic Hamiltonian \eqref{eq:H2} defines magnons, whose spectrum in the co-rotating frame is shown in Fig.~\ref{fig:fig1}(b). The dispersion is linear near $k=0$, with a minimum at the spiral wave vector $Q$, resembling the roton minimum in liquid helium. Phason at $k=0$ rotates all spins by the same angle which corresponds to a translation or a change of phase of the spiral. The magnons in the ``roton" minimum correspond to the tilt of the spiral plane, and their gap is set by the easy plane anisotropy. Collecting terms, involving three magnon operators, we get
\begin{equation}
     H^{(3)}=\sum_{\kk, \pp} \Gamma_{\kk, \pp; \kk+\pp}\alpha^\dag_{\kk} \alpha^\dag_{\pp} \alpha_{\kk+\pp}+ \mathrm{H.c.},\label{H3}
\end{equation}
with H.c. denoting Hermitian conjugate and a vertex function
\begin{equation}
\begin{split}
     \Gamma_{\kk, \pp; \kk+\pp} =&C_{\pp} u_{\kk+\pp} u_{\kk} (u_{\pp}-v_{\pp})\\
     &+C_{\kk+\pp} v_{\pp} u_{\kk} (u_{\kk+\pp}-v_{\kk+\pp})\\
     &+C_{\kk} v_{\kk+\pp} v_{\pp} (u_{\kk}-v_{\kk}),\\
     C_{\kk}=&\sum_{\delta=1,2}J_\delta \sin{k_a \delta} \sin{Q \delta}
    \end{split}
\end{equation}
We have omitted terms that create or destroy three magnons in $H^{(3)}$ because such events do not conserve energy. At low $Q$, the amplitude of scattering events involving three magnons is proportional to $Q$, therefore, we expect the thermal conductivity to decrease with increasing spiral pitch.


\paragraph{Relaxons ---}

\begin{figure*}
    \centering
\includegraphics[width=\linewidth]{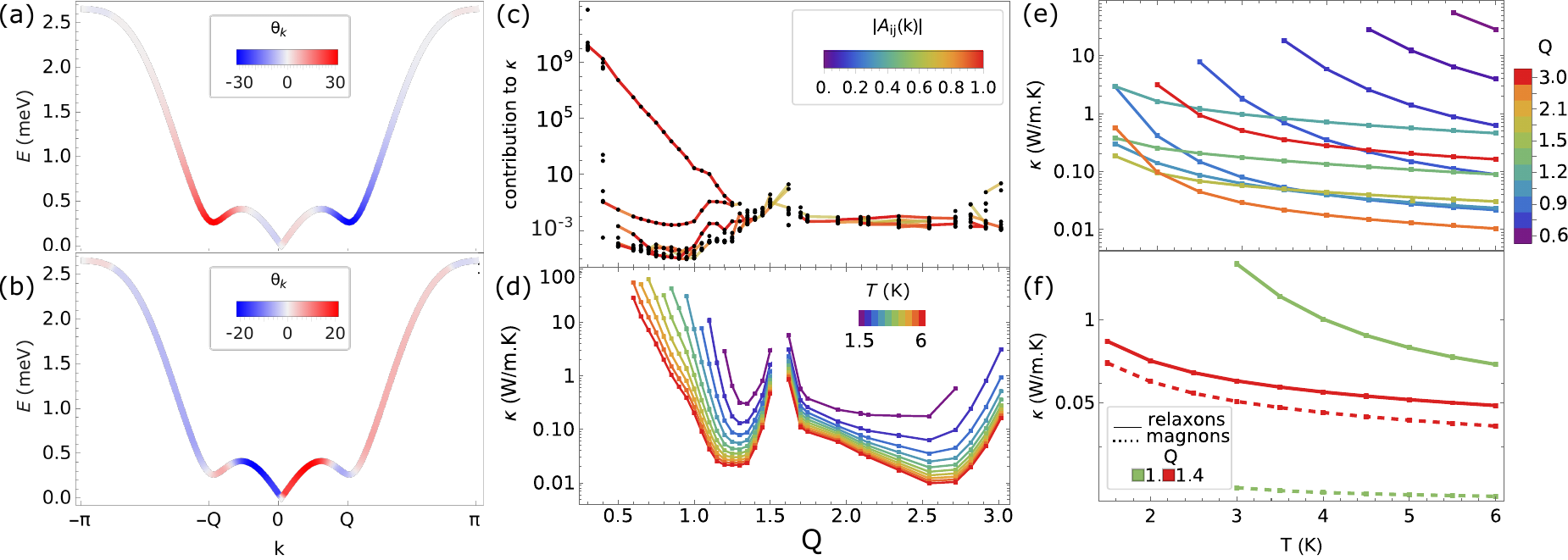}
\caption{\label{fig:relaxons}(a,b) Relaxons with the largest thermal conductivity contributions (95\% and 4\%), obtained for $T=5$~K and $Q=1.1$. The contribution $\theta_k$ of individual magnons to the relaxon is encoded by the color of the magnon dispersion curve $\omega(k)$. The positive contribution (in red) implies that the relaxon increases the magnon occupation by $\theta_k$ with respect to the Bose-Einstein equilibrium $n_k$, while blue color indicates the reduction. (c) Contributions of individual relaxons to $\kappa$ as a function of the spiral wave vector $Q$, with $J_1$ fixed and $J_2$ varied to fulfill $\cos Q=-J_1/4J_2$. (d) 
The dependence of magnon thermal conductivity on the spiral wave vector $Q$ for different temperatures. (e) Temperature dependence of magnon thermal conductivity for different spiral wave vectors $Q$. (f) Comparison between the thermal conductivity, computed in the relaxon picture and using magnon single-mode approximation.}
\end{figure*}


We employ a semiclassical treatment of thermal transport based on the evolution of the magnon distribution function, $f_{\kk}$, described by the Boltzmann transport equation with no external forces \cite{LL10},
\begin{equation}\label{eq:BTE}
    \frac{\partial f_{\kk}(\textbf{r})}{\partial t}+\textbf{v}_{\kk}\cdot \nabla f_{\kk}(\textbf{r})=\left(\frac{\partial f_{\kk}(\textbf{r})}{\partial t}\right)_{\mathrm{coll}}.
\end{equation}
The collision integral on the right hand side of the equation involves the scattering probability of a magnon $\kk$ with any other magnon. These are computed using the Fermi golden rule, with the energy and momentum conservation laws limiting the states $\kk,\qq$ of the incoming magnons to the surface, shown in Fig. \ref{fig:fig1}(c). The transition rates are then given by
\begin{multline}
        \label{beta}\beta_{\kk_1\rightarrow\kk_2\kk_3}=\frac{\pi V}{\hbar}S\left[\Gamma(\kk_1,\kk_2)+\Gamma(\kk_1,\kk_3)\right]^2\\\times\delta\left(E_{\kk_2}+E_{\kk_3}-E_{\kk_1}\right)\delta_{\kk_1, \kk_2+\kk_3}
\end{multline}        
We expand the collision integral near the equilibrium occupations, $f(k)=f_0(k)+\delta f(k)$, so that the r.h.s. takes the form $\Omega_{k,q}\delta f(k)$. Here, $\Omega_{k,q}$ is the scattering matrix, with the diagonal terms given by
\begin{widetext}
\begin{equation}\label{eq:Omegakk}
    \Omega_{\kk,\kk}=\frac{1}{NV}\sum_{\qq} \frac{1}{f_{\kk}^0(f_{\kk}^0+1)}
    \left[(P_{\kk\kk}^{\qq}-2P_{\qq\kk}^{\kk})+\sum_{\pp} 
    	\left(\frac{P_{\qq\pp}^{\kk}}{2}+P_{\kk\pp}^{\qq}\right)\right],
\end{equation}
\end{widetext}
while the off-diagonal terms are
\begin{equation}\label{eq:Omegakq}
    \Omega_{\kk\pp}=\frac{1}{NV}\sum_{\qq}\frac{1}{f_{\pp}^0(f_{\pp}^0+1)}\left(P_{\kk\pp}^{\qq}-P_{\qq\pp}^{\kk}-P_{\kk\qq}^{\pp}\right),
\end{equation}
where $N$ is the number of grid $k$-points, $V$ -- the unit cell volume, and 
\begin{equation}
    P_{\qq\pp}^{\kk}\equiv \beta_{\qq\pp}^{\kk} f_{\qq}^0 f_{\pp}^0(f_{\kk}^0+1).
\end{equation}
So far, this matrix is nonsymmetric due to factors in the denominator. To diagonalize it, we first perform the transformation \cite{Hardy63,Hardy70}
\begin{equation}\label{eq:trasfsymm}
\begin{split}
    &\Tilde{\Omega}_{\kk\pp}=\Omega_{\kk\pp}\sqrt{\frac{f^0_{\pp}(f^0_{\pp}+1)}{f^0_{\kk}(f^0_{\kk}+1)}}\\
    &\Tilde{f}_{\kk}=\frac{f_{\kk}}{\sqrt{f_{\kk}^0(f_{\kk}^0+1)}}  
    \end{split}
\end{equation}
The new matrix $\Tilde{\Omega}_{\kk\pp}$ is symmetric. 
Since it is not diagonal, individual magnons are not the eigenvectors and thus are not heat carriers -- a single magnon does not just decay exponentially with a well-defined relaxation time, but instead scatters into other magnon states. The collision matrix can be diagonalized and its eigenvectors $\theta_\mu$ describe actual heat carriers – relaxons, which are linear combinations of many magnons \cite{Cepellotti16}. Inverse eigenvalues give the relaxation times, $\tau_\mu$.

\paragraph{Results ---} We implemented a code \cite{MargCond} that computes the scattering matrix and diagonalizes it to obtain relaxons and thermal conductivity. The reciprocal space is discretized on a grid, and the convergence of thermal conductivity with respect to the grid density is achieved on a $128\times 8\times 8$ $k$-grid 
, with 128 $k$-points along the spiral wave vector.

Fig.~\ref{fig:relaxons} shows the relaxons with the highest contributions to $\kappa$ obtained for $T=5$~K and $Q=1.1$~rad. The line represents the magnon dispersion, while its opacity and color (positive in red, negative in blue) encode the modulation of magnon occupation numbers $\theta_k$ defining the relaxon. Here, $\theta_k$ is odd, $\theta(k)=-\theta(-k)$, as is the magnon velocity $v_k=\partial\omega_k/\partial k$, leading to non-zero heat flux. The relaxon in Fig.~\ref{fig:relaxons}(a) contributes 95\% to $\kappa$ and is dominated by magnons at the ``roton'' minimum $k_x\approx\pm Q$. They correspond to the oscillations of the spiral plane, and, in the case of cycloidal spiral ground state, are Katsura-Balatsky-Nagaosa (KBN) electromagnons \cite{Katsura2005} that induce oscillations of the ferroelectric polarization. The relaxon in Fig.~\ref{fig:relaxons}(b) is responsible for 4\% of $\kappa$ and is dominated by acoustic magnons. $\theta_k$ here describes extra magnons with the positive velocity component added to the equilibrium magnon distribution, and a reduced number of magnons traveling in the opposite direction, giving a net heat flux.
Fig. \ref{fig:relaxons}(c) shows the contributions of individual relaxons at $T=2.5$ K. In spiral magnets only a few relaxons significantly contribute to the thermal conductivity, as is also found in phononic systems of recent interest. 
The lines connect relaxons of similar character, with their intensity computed as an overlap between relaxons at neighboring $k$-points, $A_{ij}(k)=\langle \delta n_i(Q)|\delta n_j(Q+\delta Q)\rangle$, \cite{MV1997}. At small spiral wave vectors $Q\ll 1$, and in fact all the way to $Q\sim 1$, a single relaxon shown in Fig.~\ref{fig:relaxons}(a) dominates, while all others have a negligible contribution. From $Q=1$, another relaxon, shown in Fig.~\ref{fig:relaxons}(b), starts to contribute.

The dependence of the thermal conductivity on the spiral wave vector $Q$ and temperature is shown in Fig.~\ref{fig:relaxons}(d,e), where we fix $J_2=1$~meV and vary $J_1=-4J_2\cos Q$. At low temperatures $k_{\mathrm{B}}T\ll J_1$, $\kappa$ increases exponentially with decreasing $Q$, as seen in Fig.~\ref{fig:relaxons}(d). This can be understood as follows: $\Gamma$ includes normal ($k,q\to k+q$) and Umklapp processes $k,q\to k+q+G$, where $G$ is a reciprocal lattice vector. Normal processes do not change the total momentum of magnons, and therefore they cannot produce a heat flux in response to a temperature gradient. Umklapp processes do change the total momentum of magnons. However, in order to satisfy the momentum conservation law in the Umklapp process, at least one of the incoming magnons must have a wave vector of the order of $G$. Such magnons have a high energy of the order of $J_1,J_2$ and thus low occupation numbers $n_k\approx \exp(-\epsilon_k/T)$ at low $T$. For larger spiral wave vectors, $Q\approx 2\pi/3$, an Umklapp process involving magnons at roton minima becomes possible, leading to a decrease in thermal conductivity. 
The exponential increase of the thermal conductivity is limited by the onset of a regime, in which Umklapp scattering does not occur within the sample size. The kinetic equation approach becomes invalid when the relaxon mean free path reaches the sample size, and the energy transport becomes ballistic. This corresponds to the end point of the curves in Figs.~\ref{fig:relaxons}(d,e). Past that point, the Landauer-Buttiker formalism must be used, and quantized thermal conductance is expected. For FM spirals, $\kappa$ reaches its minimum at 1.3 rad$\approx 70^\circ$.
As $Q$ approaches $\pi/2$, $J_1$ goes to zero, thus decoupling the subsystems made of odd and even sites from each other and creating two independent AFM systems. Their collinearity implies the decrease of three-magnon scattering and an increase of $\kappa$ towards $Q=\pi/2$.
When $Q$ is increased past $\pi/2$, into an AFM spiral region, $\kappa$ reaches a minimum at around 2.6 rad=150$^\circ$ and then increases towards a collinear AFM state at $Q=\pi$.

In principle, the relaxons and $\kappa$ depend on the energy smearing $\sigma_E$ of the Gaussian, used to approximate the delta-function in energy in Eq.~\ref{beta}. In Fig.~\ref{fig:relaxons}(a-f) we used $\sigma_E=E(Q/2)/5$ (see SI for details). 

Comparing the values of thermal conductivity estimated from magnon single-mode and relaxon pictures, shown in Fig.~\ref{fig:relaxons}(f), we see that the magnon picture always gives lower thermal conductivity. That is because the relaxation time for a magnon in a single-mode approximation corresponds to the scattering to the state with another wave vector, which does not necessarily cause energy flux dissipation and may also contribute to conducting heat in the same direction. At the same time, a relaxon describes a combination of many magnon states, that scatter into each other, together having a purely exponential decay without scattering to other relaxon states (to the considered order of expansion in \eqref{H3}), therefore the relaxation time of the relaxon that contributes the most to $\kappa$ is higher than that of a magnon in SMA.



$\kappa_{\mathrm{mag}}$ computed in a single mode approximation decreases with temperature at high $Q$, but increases at low $Q$, even though the magnon relaxation rates increase with $T$. This peculiar behavior originates from the flat dispersion near the center of the zone at low $Q$, leading to a low magnon velocity $v$ in $\kappa_k\propto v_k^2\tau_k$. At higher $T$, magnon states with higher velocity are occupied, as seen in Fig.~\ref{fig:fig1}(b), therefore driving $\kappa_{\mathrm{mag}}$ higher.

\paragraph{Conclusions ---}
We presented the relaxon theory of thermal conductivity for spiral magnets. A single relaxon composed of KBN magnons representing the oscillations of the spiral plane accounts for the main contribution to $\kappa$ in a wide range of spiral wave vectors and temperatures. The next-largest contribution is given by the relaxon made of acoustic magnons. Thermal conductivity increases exponentially at low $T$, and at low spiral wave vector. Noncollinearity reduces thermal conductivity, consistent with experiments on GaV$_4$S$_8$ where switching between ferromagnetic, spiral, and skyrmion lattice states was achieved under small magnetic fields  and a 50\% decrease of $\kappa$ was observed in noncollinear states \cite{slowdown}. Skyrmion tubes may therefore act as thermal conductors, guiding heat currents along the core. Since KBN electromagnons constitute the dominant relaxon, and their gap may be controlled with magnetic and electric fields, heat valve functionality may be envisaged. The results open the path to exploring effects of spin texture topology, anomalous magnon velocity, and magnon viscosity on thermal transport.




\paragraph{Acknowledgements ---}
Fruitful discussions with N. Nagaosa, N. Marzari, M. Mostovoy and F. Foggetti are gratefully acknowledged.


\begin{thebibliography}{21}%
\makeatletter
\providecommand \@ifxundefined [1]{%
 \@ifx{#1\undefined}
}%
\providecommand \@ifnum [1]{%
 \ifnum #1\expandafter \@firstoftwo
 \else \expandafter \@secondoftwo
 \fi
}%
\providecommand \@ifx [1]{%
 \ifx #1\expandafter \@firstoftwo
 \else \expandafter \@secondoftwo
 \fi
}%
\providecommand \natexlab [1]{#1}%
\providecommand \enquote  [1]{``#1''}%
\providecommand \bibnamefont  [1]{#1}%
\providecommand \bibfnamefont [1]{#1}%
\providecommand \citenamefont [1]{#1}%
\providecommand \href@noop [0]{\@secondoftwo}%
\providecommand \href [0]{\begingroup \@sanitize@url \@href}%
\providecommand \@href[1]{\@@startlink{#1}\@@href}%
\providecommand \@@href[1]{\endgroup#1\@@endlink}%
\providecommand \@sanitize@url [0]{\catcode `\\12\catcode `\$12\catcode
  `\&12\catcode `\#12\catcode `\^12\catcode `\_12\catcode `\%12\relax}%
\providecommand \@@startlink[1]{}%
\providecommand \@@endlink[0]{}%
\providecommand \url  [0]{\begingroup\@sanitize@url \@url }%
\providecommand \@url [1]{\endgroup\@href {#1}{\urlprefix }}%
\providecommand \urlprefix  [0]{URL }%
\providecommand \Eprint [0]{\href }%
\providecommand \doibase [0]{https://doi.org/}%
\providecommand \selectlanguage [0]{\@gobble}%
\providecommand \bibinfo  [0]{\@secondoftwo}%
\providecommand \bibfield  [0]{\@secondoftwo}%
\providecommand \translation [1]{[#1]}%
\providecommand \BibitemOpen [0]{}%
\providecommand \bibitemStop [0]{}%
\providecommand \bibitemNoStop [0]{.\EOS\space}%
\providecommand \EOS [0]{\spacefactor3000\relax}%
\providecommand \BibitemShut  [1]{\csname bibitem#1\endcsname}%
\let\auto@bib@innerbib\@empty
\bibitem [{\citenamefont {Katsura}\ \emph {et~al.}(2005)\citenamefont
  {Katsura}, \citenamefont {Nagaosa},\ and\ \citenamefont
  {Balatsky}}]{Katsura2005}%
  \BibitemOpen
  \bibfield  {author} {\bibinfo {author} {\bibfnamefont {H.}~\bibnamefont
  {Katsura}}, \bibinfo {author} {\bibfnamefont {N.}~\bibnamefont {Nagaosa}},\
  and\ \bibinfo {author} {\bibfnamefont {A.~V.}\ \bibnamefont {Balatsky}},\
  }\bibfield  {title} {\bibinfo {title} {Spin current and magnetoelectric
  effect in noncollinear magnets},\ }\href
  {https://doi.org/10.1103/PhysRevLett.95.057205} {\bibfield  {journal}
  {\bibinfo  {journal} {Phys. Rev. Lett.}\ }\textbf {\bibinfo {volume} {95}},\
  \bibinfo {pages} {057205} (\bibinfo {year} {2005})}\BibitemShut {NoStop}%
\bibitem [{\citenamefont {Mostovoy}(2006)}]{Mostovoy2006}%
  \BibitemOpen
  \bibfield  {author} {\bibinfo {author} {\bibfnamefont {M.}~\bibnamefont
  {Mostovoy}},\ }\bibfield  {title} {\bibinfo {title} {Ferroelectricity in
  spiral magnets},\ }\href {https://doi.org/10.1103/PhysRevLett.96.067601}
  {\bibfield  {journal} {\bibinfo  {journal} {Phys. Rev. Lett.}\ }\textbf
  {\bibinfo {volume} {96}},\ \bibinfo {pages} {067601} (\bibinfo {year}
  {2006})}\BibitemShut {NoStop}%
\bibitem [{\citenamefont {Sergienko}\ and\ \citenamefont
  {Dagotto}(2006)}]{Sergienko2006}%
  \BibitemOpen
  \bibfield  {author} {\bibinfo {author} {\bibfnamefont {I.~A.}\ \bibnamefont
  {Sergienko}}\ and\ \bibinfo {author} {\bibfnamefont {E.}~\bibnamefont
  {Dagotto}},\ }\bibfield  {title} {\bibinfo {title} {Role of the
  {D}zyaloshinskii-{M}oriya interaction in multiferroic perovskites},\ }\href
  {https://doi.org/10.1103/PhysRevB.73.094434} {\bibfield  {journal} {\bibinfo
  {journal} {Phys. Rev. B}\ }\textbf {\bibinfo {volume} {73}},\ \bibinfo
  {pages} {094434} (\bibinfo {year} {2006})}\BibitemShut {NoStop}%
\bibitem [{\citenamefont {Cheong}\ and\ \citenamefont
  {Mostovoy}(2007)}]{cheong2007multiferroics}%
  \BibitemOpen
  \bibfield  {author} {\bibinfo {author} {\bibfnamefont {S.-W.}\ \bibnamefont
  {Cheong}}\ and\ \bibinfo {author} {\bibfnamefont {M.}~\bibnamefont
  {Mostovoy}},\ }\bibfield  {title} {\bibinfo {title} {Multiferroics: a
  magnetic twist for ferroelectricity},\ }\href@noop {} {\bibfield  {journal}
  {\bibinfo  {journal} {Nature Materials}\ }\textbf {\bibinfo {volume} {6}},\
  \bibinfo {pages} {13} (\bibinfo {year} {2007})}\BibitemShut {NoStop}%
\bibitem [{\citenamefont {Psaroudaki}\ and\ \citenamefont
  {Panagopoulos}(2021)}]{Psaroudaki21}%
  \BibitemOpen
  \bibfield  {author} {\bibinfo {author} {\bibfnamefont {C.}~\bibnamefont
  {Psaroudaki}}\ and\ \bibinfo {author} {\bibfnamefont {C.}~\bibnamefont
  {Panagopoulos}},\ }\bibfield  {title} {\bibinfo {title} {Skyrmion qubits: A
  new class of quantum logic elements based on nanoscale magnetization},\
  }\href {https://doi.org/10.1103/PhysRevLett.127.067201} {\bibfield  {journal}
  {\bibinfo  {journal} {Phys. Rev. Lett.}\ }\textbf {\bibinfo {volume} {127}},\
  \bibinfo {pages} {067201} (\bibinfo {year} {2021})}\BibitemShut {NoStop}%
\bibitem [{\citenamefont {Psaroudaki}\ \emph {et~al.}(2023)\citenamefont
  {Psaroudaki}, \citenamefont {Peraticos},\ and\ \citenamefont
  {Panagopoulos}}]{Psaroudaki23}%
  \BibitemOpen
  \bibfield  {author} {\bibinfo {author} {\bibfnamefont {C.}~\bibnamefont
  {Psaroudaki}}, \bibinfo {author} {\bibfnamefont {E.}~\bibnamefont
  {Peraticos}},\ and\ \bibinfo {author} {\bibfnamefont {C.}~\bibnamefont
  {Panagopoulos}},\ }\bibfield  {title} {\bibinfo {title} {{Skyrmion qubits:
  Challenges for future quantum computing applications}},\ }\href
  {https://doi.org/10.1063/5.0177864} {\bibfield  {journal} {\bibinfo
  {journal} {Applied Physics Letters}\ }\textbf {\bibinfo {volume} {123}},\
  \bibinfo {pages} {260501} (\bibinfo {year} {2023})},\ \Eprint
  {https://arxiv.org/abs/https://pubs.aip.org/aip/apl/article-pdf/doi/10.1063/5.0177864/18274001/260501\_1\_5.0177864.pdf}
  {https://pubs.aip.org/aip/apl/article-pdf/doi/10.1063/5.0177864/18274001/260501\_1\_5.0177864.pdf}
  \BibitemShut {NoStop}%
\bibitem [{\citenamefont {Qu}\ \emph {et~al.}(2025)\citenamefont {Qu},
  \citenamefont {Zou}, \citenamefont {Loss},\ and\ \citenamefont
  {Hirosawa}}]{Loss25}%
  \BibitemOpen
  \bibfield  {author} {\bibinfo {author} {\bibfnamefont {G.}~\bibnamefont
  {Qu}}, \bibinfo {author} {\bibfnamefont {J.}~\bibnamefont {Zou}}, \bibinfo
  {author} {\bibfnamefont {D.}~\bibnamefont {Loss}},\ and\ \bibinfo {author}
  {\bibfnamefont {T.}~\bibnamefont {Hirosawa}},\ }\bibfield  {title} {\bibinfo
  {title} {Density matrix renormalization group study of domain wall qubits},\
  }\href {https://doi.org/10.1103/mpgp-pqk2} {\bibfield  {journal} {\bibinfo
  {journal} {Phys. Rev. B}\ }\textbf {\bibinfo {volume} {112}},\ \bibinfo
  {pages} {054432} (\bibinfo {year} {2025})}\BibitemShut {NoStop}%
\bibitem [{\citenamefont {Sekiguchi}\ \emph {et~al.}(2022)\citenamefont
  {Sekiguchi}, \citenamefont {Budzinauskas}, \citenamefont {Padmanabhan},
  \citenamefont {Versteeg}, \citenamefont {Tsurkan}, \citenamefont
  {K{\'e}zsm{\'a}rki}, \citenamefont {Foggetti}, \citenamefont {Artyukhin},\
  and\ \citenamefont {van Loosdrecht}}]{slowdown}%
  \BibitemOpen
  \bibfield  {author} {\bibinfo {author} {\bibfnamefont {F.}~\bibnamefont
  {Sekiguchi}}, \bibinfo {author} {\bibfnamefont {K.}~\bibnamefont
  {Budzinauskas}}, \bibinfo {author} {\bibfnamefont {P.}~\bibnamefont
  {Padmanabhan}}, \bibinfo {author} {\bibfnamefont {R.~B.}\ \bibnamefont
  {Versteeg}}, \bibinfo {author} {\bibfnamefont {V.}~\bibnamefont {Tsurkan}},
  \bibinfo {author} {\bibfnamefont {I.}~\bibnamefont {K{\'e}zsm{\'a}rki}},
  \bibinfo {author} {\bibfnamefont {F.}~\bibnamefont {Foggetti}}, \bibinfo
  {author} {\bibfnamefont {S.}~\bibnamefont {Artyukhin}},\ and\ \bibinfo
  {author} {\bibfnamefont {P.~H.}\ \bibnamefont {van Loosdrecht}},\ }\bibfield
  {title} {\bibinfo {title} {Slowdown of photoexcited spin dynamics in the
  non-collinear spin-ordered phases in skyrmion host {GaV4S8}},\ }\href@noop {}
  {\bibfield  {journal} {\bibinfo  {journal} {Nature Communications}\ }\textbf
  {\bibinfo {volume} {13}},\ \bibinfo {pages} {1} (\bibinfo {year}
  {2022})}\BibitemShut {NoStop}%
\bibitem [{\citenamefont {Parodi}\ \emph {et~al.}()\citenamefont {Parodi},
  \citenamefont {Parrinello},\ and\ \citenamefont
  {S.Artyukhin}}]{ParodiGenetic}%
  \BibitemOpen
  \bibfield  {author} {\bibinfo {author} {\bibfnamefont {M.}~\bibnamefont
  {Parodi}}, \bibinfo {author} {\bibfnamefont {M.}~\bibnamefont {Parrinello}},\
  and\ \bibinfo {author} {\bibnamefont {S.Artyukhin}},\ }\href@noop {}
  {\bibinfo {title} {Genetic algorithm for spin textures}}\BibitemShut
  {NoStop}%
\bibitem [{\citenamefont {Hardy}(1970)}]{Hardy70}%
  \BibitemOpen
  \bibfield  {author} {\bibinfo {author} {\bibfnamefont {R.~J.}\ \bibnamefont
  {Hardy}},\ }\bibfield  {title} {\bibinfo {title} {Phonon {Boltzmann} equation
  and second sound in solids},\ }\href
  {https://doi.org/10.1103/PhysRevB.2.1193} {\bibfield  {journal} {\bibinfo
  {journal} {Phys. Rev. B}\ }\textbf {\bibinfo {volume} {2}},\ \bibinfo {pages}
  {1193} (\bibinfo {year} {1970})}\BibitemShut {NoStop}%
\bibitem [{\citenamefont {Cepellotti}\ and\ \citenamefont
  {Marzari}(2016)}]{Cepellotti16}%
  \BibitemOpen
  \bibfield  {author} {\bibinfo {author} {\bibfnamefont {A.}~\bibnamefont
  {Cepellotti}}\ and\ \bibinfo {author} {\bibfnamefont {N.}~\bibnamefont
  {Marzari}},\ }\bibfield  {title} {\bibinfo {title} {Thermal transport in
  crystals as a kinetic theory of relaxons},\ }\href
  {https://doi.org/10.1103/PhysRevX.6.041013} {\bibfield  {journal} {\bibinfo
  {journal} {Phys. Rev. X}\ }\textbf {\bibinfo {volume} {6}},\ \bibinfo {pages}
  {041013} (\bibinfo {year} {2016})}\BibitemShut {NoStop}%
\bibitem [{\citenamefont {Manipatruni}\ \emph {et~al.}(2018)\citenamefont
  {Manipatruni}, \citenamefont {Nikonov},\ and\ \citenamefont
  {Young}}]{Manipatruni2018}%
  \BibitemOpen
  \bibfield  {author} {\bibinfo {author} {\bibfnamefont {S.}~\bibnamefont
  {Manipatruni}}, \bibinfo {author} {\bibfnamefont {D.~E.}\ \bibnamefont
  {Nikonov}},\ and\ \bibinfo {author} {\bibfnamefont {I.~A.}\ \bibnamefont
  {Young}},\ }\bibfield  {title} {\bibinfo {title} {Beyond {CMOS} computing
  with spin and polarization},\ }\href
  {https://doi.org/10.1038/s41567-018-0101-4} {\bibfield  {journal} {\bibinfo
  {journal} {Nature Physics}\ }\textbf {\bibinfo {volume} {14}},\ \bibinfo
  {pages} {338} (\bibinfo {year} {2018})}\BibitemShut {NoStop}%
\bibitem [{\citenamefont {Dmitriev}\ and\ \citenamefont
  {Krivnov}(2006)}]{Dmitriev06}%
  \BibitemOpen
  \bibfield  {author} {\bibinfo {author} {\bibfnamefont {D.~V.}\ \bibnamefont
  {Dmitriev}}\ and\ \bibinfo {author} {\bibfnamefont {V.~Y.}\ \bibnamefont
  {Krivnov}},\ }\bibfield  {title} {\bibinfo {title} {Frustrated ferromagnetic
  spin-$\frac{1}{2}$ chain in a magnetic field},\ }\href
  {https://doi.org/10.1103/PhysRevB.73.024402} {\bibfield  {journal} {\bibinfo
  {journal} {Phys. Rev. B}\ }\textbf {\bibinfo {volume} {73}},\ \bibinfo
  {pages} {024402} (\bibinfo {year} {2006})}\BibitemShut {NoStop}%
\bibitem [{\citenamefont {Holstein}\ and\ \citenamefont
  {Primakoff}(1940)}]{HP40}%
  \BibitemOpen
  \bibfield  {author} {\bibinfo {author} {\bibfnamefont {T.}~\bibnamefont
  {Holstein}}\ and\ \bibinfo {author} {\bibfnamefont {H.}~\bibnamefont
  {Primakoff}},\ }\bibfield  {title} {\bibinfo {title} {Field dependence of the
  intrinsic domain magnetization of a ferromagnet},\ }\href
  {https://doi.org/10.1103/PhysRev.58.1098} {\bibfield  {journal} {\bibinfo
  {journal} {Phys. Rev.}\ }\textbf {\bibinfo {volume} {58}},\ \bibinfo {pages}
  {1098} (\bibinfo {year} {1940})}\BibitemShut {NoStop}%
\bibitem [{\citenamefont {Bogolyubov}\ \emph {et~al.}(1958)\citenamefont
  {Bogolyubov}, \citenamefont {Tolmachev},\ and\ \citenamefont
  {Shirkov}}]{Bogolyubov1958}%
  \BibitemOpen
  \bibfield  {author} {\bibinfo {author} {\bibfnamefont {N.~N.}\ \bibnamefont
  {Bogolyubov}}, \bibinfo {author} {\bibfnamefont {V.~V.}\ \bibnamefont
  {Tolmachev}},\ and\ \bibinfo {author} {\bibfnamefont {D.~V.}\ \bibnamefont
  {Shirkov}},\ }\bibfield  {title} {\bibinfo {title} {{A New method in the
  theory of superconductivity}},\ }\href
  {https://doi.org/10.1002/prop.19580061102} {\bibfield  {journal} {\bibinfo
  {journal} {Fortsch. Phys.}\ }\textbf {\bibinfo {volume} {6}},\ \bibinfo
  {pages} {605} (\bibinfo {year} {1958})}\BibitemShut {NoStop}%
\bibitem [{\citenamefont {Valatin}(1958)}]{Valatin1958}%
  \BibitemOpen
  \bibfield  {author} {\bibinfo {author} {\bibfnamefont {J.~G.}\ \bibnamefont
  {Valatin}},\ }\bibfield  {title} {\bibinfo {title} {{Comments on the theory
  of superconductivity}},\ }\href {https://doi.org/10.1007/BF02745589}
  {\bibfield  {journal} {\bibinfo  {journal} {Il Nuovo Cimento}\ }\textbf
  {\bibinfo {volume} {7}},\ \bibinfo {pages} {843–857} (\bibinfo {year}
  {1958})}\BibitemShut {NoStop}%
\bibitem [{\citenamefont {Lifshitz}\ and\ \citenamefont
  {Pitaevskii}(1995)}]{LL10}%
  \BibitemOpen
  \bibfield  {author} {\bibinfo {author} {\bibfnamefont {E.}~\bibnamefont
  {Lifshitz}}\ and\ \bibinfo {author} {\bibfnamefont {L.}~\bibnamefont
  {Pitaevskii}},\ }\href {https://books.google.it/books?id=HgRBAQAAIAAJ} {\emph
  {\bibinfo {title} {Physical Kinetics: Volume 10}}},\ Course of theoretical
  physics\ (\bibinfo  {publisher} {Elsevier Science},\ \bibinfo {year}
  {1995})\BibitemShut {NoStop}%
\bibitem [{\citenamefont {Hardy}(1963)}]{Hardy63}%
  \BibitemOpen
  \bibfield  {author} {\bibinfo {author} {\bibfnamefont {R.~J.}\ \bibnamefont
  {Hardy}},\ }\bibfield  {title} {\bibinfo {title} {Energy-flux operator for a
  lattice},\ }\href {https://doi.org/10.1103/PhysRev.132.168} {\bibfield
  {journal} {\bibinfo  {journal} {Phys. Rev.}\ }\textbf {\bibinfo {volume}
  {132}},\ \bibinfo {pages} {168} (\bibinfo {year} {1963})}\BibitemShut
  {NoStop}%
\bibitem [{\citenamefont {Parodi}(2023)}]{MargCond}%
  \BibitemOpen
  \bibfield  {author} {\bibinfo {author} {\bibfnamefont {M.}~\bibnamefont
  {Parodi}},\ }\href@noop {} {\bibinfo {title} {Margcond}},\ \bibinfo
  {howpublished} {\url{https://github.com/SergeyArtyukhin/viscosity}} (\bibinfo
  {year} {2023}),\ \bibinfo {note} {commit available at
  \url{https://github.com/SergeyArtyukhin/viscosity}}\BibitemShut {NoStop}%
\bibitem [{\citenamefont {Marzari}\ and\ \citenamefont
  {Vanderbilt}(1997)}]{MV1997}%
  \BibitemOpen
  \bibfield  {author} {\bibinfo {author} {\bibfnamefont {N.}~\bibnamefont
  {Marzari}}\ and\ \bibinfo {author} {\bibfnamefont {D.}~\bibnamefont
  {Vanderbilt}},\ }\bibfield  {title} {\bibinfo {title} {Maximally localized
  generalized {Wannier} functions for composite energy bands},\ }\href
  {https://doi.org/10.1103/PhysRevB.56.12847} {\bibfield  {journal} {\bibinfo
  {journal} {Phys. Rev. B}\ }\textbf {\bibinfo {volume} {56}},\ \bibinfo
  {pages} {12847} (\bibinfo {year} {1997})}\BibitemShut {NoStop}%
\bibitem [{\citenamefont {Du}\ \emph {et~al.}(2016)\citenamefont {Du},
  \citenamefont {Liu}, \citenamefont {Xie}, \citenamefont {Wang},\ and\
  \citenamefont {Liu}}]{chinese16}%
  \BibitemOpen
  \bibfield  {author} {\bibinfo {author} {\bibfnamefont {Z.~Z.}\ \bibnamefont
  {Du}}, \bibinfo {author} {\bibfnamefont {H.~M.}\ \bibnamefont {Liu}},
  \bibinfo {author} {\bibfnamefont {Y.~L.}\ \bibnamefont {Xie}}, \bibinfo
  {author} {\bibfnamefont {Q.~H.}\ \bibnamefont {Wang}},\ and\ \bibinfo
  {author} {\bibfnamefont {J.-M.}\ \bibnamefont {Liu}},\ }\bibfield  {title}
  {\bibinfo {title} {Magnetic excitations in quasi-one-dimensional helimagnets:
  Magnon decays and influence of interchain interactions},\ }\href
  {https://doi.org/10.1103/PhysRevB.94.134416} {\bibfield  {journal} {\bibinfo
  {journal} {Phys. Rev. B}\ }\textbf {\bibinfo {volume} {94}},\ \bibinfo
  {pages} {134416} (\bibinfo {year} {2016})}\BibitemShut {NoStop}%
\end{thebibliography}
%

\widetext
\newpage
\setcounter{equation}{0}
\setcounter{figure}{0}
\setcounter{table}{0}

\begin{center}

\textbf{\large SUPPLEMENTARY MATERIAL}
\end{center}

{\bf Holstein-Primakoff transformation for a spin spiral}
\\
In order to describe the magnetic excitations in our system, we have to apply the Holstein-Primakoff transformation \cite{HP40} to the Hamiltonian Eq.~(\ref{eq:Ham}). However, simply doing so would mean that we are expanding around the ferromagnetic state $S_z=S$, while the expansion must be performed around the classical directions of the spins in the spiral. Therefore, we apply a rotation around the $y$ axis, $\hat R_i\equiv\hat R(\theta_i)$, where $\theta_i$ is the deviation of the $i$-th spin from the $z$ axis, to align the state with zero bosons with a classical spin direction on site $i$. The rotation matrix is then 
\begin{equation}\label{eq:rotationmat}
    \hat R_\ii=\hat R(\theta_\ii)=\begin{pmatrix}
    \cos{\theta_\ii}&0&-\sin{\theta_\ii}\\
    0&1&0\\
    \sin{\theta_\ii}&0&\cos{\theta_\ii}\\
    \end{pmatrix}.
\end{equation}
The spin on site $\ii$ is now $\hat R_\ii\mathbf{S}_\ii$ and the scalar product of the two spins
\begin{equation}\label{eq:SRS}
    (\hat R_\ii\mathbf{S}_\ii)\cdot\hat R_\jj\mathbf{S}_\jj=(\hat R_\ii\mathbf{S}_\ii)^T\hat R_\jj\mathbf{S}_\jj=\mathbf{S}_\ii^T\hat R_\ii^T\hat R_\jj\mathbf{S}_\jj=\mathbf{S}_\ii^T\hat R(\theta_\jj-\theta_\ii)\mathbf{S}_\jj.
\end{equation}
Then the Hamiltonian takes the form:
\begin{equation}\label{eq:rotatedHamxyz}
     H= \sum_{\ii\jj} J_{\ii\jj} \left[S^y_\ii S^y_\jj + \cos{\theta_{\ii\jj}}\left( S^x_\ii S^x_\jj +S^z_\ii s^z_\jj \right)+\sin{\theta_{\ii\jj}}\left(S^x_\ii S^z_\jj - S^z_\ii S^x_\jj \right)\right]+\Delta \left(S_\ii^y\right)^2
\end{equation}
where $\theta_{\ii\jj}\equiv\theta_\jj-\theta_\ii$ takes the values 
\begin{equation}
    \theta_{\ii\jj}=\begin{cases}
    Q, & \text{if}\ \jj=\ii \pm \textbf{a} \\
    2Q, & \text{if}\ \jj=\ii \pm 2\textbf{a} \\
    0, & \text{otherwise}\ 
    \end{cases}
\end{equation}
and 
\begin{equation}
    J_{\ii\jj}=\begin{cases}
    J_1, & \text{if}\ \jj=\ii \pm \textbf{a} \\
    J_2, & \text{if}\ \jj=\ii \pm 2\textbf{a} \\
    J_3, & \text{if}\ \jj=\ii \pm \textbf{b} \\
    J_4, & \text{if}\ \jj=\ii \pm \textbf{c} \\
    0, & \text{otherwise}\ 
    \end{cases}
\end{equation}

{\bf Third-order Hamiltonian}
\\
To derive the three magnon Hamiltonian, we note that the Holstein-Primakoff transformation \cite{HP40}, with the square root expanded up to the first order in $1/S$, reads
\begin{equation}
    \begin{split}
        &S^x_{\ii}=\frac{\sqrt{2S}}{2}\left[\left(a_{\ii}-\frac{a^\dag_{\ii} a_{\ii} a_{\ii}}{4S}\right)+\left(a^\dag_{\ii}-\frac{a^\dag_{\ii} a^\dag_{\ii} a_{\ii}}{4S}\right)\right],\\
        &S^y_{\ii}=\frac{\sqrt{2S}}{2i}\left[\left(a_{\ii}-\frac{a^\dag_{\ii} a_{\ii} a_{\ii}}{4S}\right)-\left(a^\dag_{\ii}-\frac{a^\dag_{\ii} a^\dag_{\ii} a_{\ii}}{4S}\right)\right],\\
        &S^z_{\ii}= S - a^\dag_{\ii} a_{\ii}.\\ 
    \end{split}
\end{equation}
Then, it is clear that, from terms proportional to $S^xS^x$, $S^yS^y$, and $S^zS^z$, we only get processes involving at least four magnons. However, in the non-collinear case \cite{Dmitriev06,chinese16}, the presence of the rotation matrix in the scalar product between the spins, Eq.~(\ref{eq:SRS}), gives rise to terms involving $S^xS^z$ (and $S^zS^x$) and $S^yS^z$ (and $S^zS^y$), as can be seen in Eq.~(\ref{eq:rotatedHamxyz}).
We perform the Holstein-Primakoff transformation in the Hamiltonian Eq.~(\ref{eq:rotatedHamxyz}). Collecting terms involving three magnons, we are left with 
\begin{equation}
        H^{(3)}=\frac{\sqrt{2S}}{2}\sum_{\ii\jj} J_{\ii\jj}\sin{\theta_{\ii\jj}}\left[a^\dag_\ii a_\ii\left(a_\jj+a^\dag_\jj\right)-\left(a_\ii+a^\dag_\ii\right)a^\dag_\jj a_\jj\right].
\end{equation}
We proceed with the Fourier transform:
\begin{equation}\label{eq:FourierTransform}
    a_\ii=\frac{1}{\sqrt{N}}\sum_\kk a_\kk e^{i\kk\cdot \textbf{R}_i}
\end{equation}
to get
\begin{equation}
    \begin{split}
        H^{(3)}= \frac{1}{N^{3/2}}\frac{\sqrt{2S}}{2}\sum_{\delta=1,2} \sum_{\ii} J_\delta \sin{(\delta Q)}\sum_{\kk\qq\pp}&a^\dag_\kk a_\qq e^{i(\qq-\kk)\cdot\RR_i}\left[e^{i(q_x-k_x)\delta a}\left(a_\pp e^{i\pp\cdot\RR_i}+a^\dag_\pp e^{-i\pp\cdot\RR_i}\right)\right.\\
        &\left.+\left(a_\pp e^{i\pp\cdot\RR_i}e^{ip_x \delta a}+a^\dag_\pp e^{-i\pp\cdot\RR_i}e^{-ip_x \delta a}\right)\right].
    \end{split}
\end{equation}
Here, we used the fact that $\sin{\theta_{\ii\jj}}=0$ for $\jj=\ii+\textbf{b}$ and $\jj=\ii+\textbf{c}$.
Using the relation $\sum_{\ii}e^{i(\kk-\qq)\cdot\RR_i}=N\delta_{\kk,\qq}$, we can sum over $\ii$:
\begin{equation}
    \begin{split}
       H^{(3)}=  \frac{\sqrt{2S}}{\sqrt{N}} i \sum_{\delta} J_{\delta}\sin{(\delta Q)}\sum_{\kk\qq}\sin{(\delta k_x a)}\left(a^\dag_\kk a_{\kk+\qq}a^\dag_\qq-a^\dag_{\kk+\qq} a_\qq a_\kk\right)
    \end{split}
\end{equation}
The final Hamiltonian is \cite{Dmitriev06}
\begin{equation}
    H^{(3)}=\frac{\sqrt{2S}}{2\sqrt{N}}i\sum_{\kk\qq}\left(\beta(k_x)+\beta(q_x)\right)\left(a^\dag_\kk a_{\kk+\qq}a^\dag_\qq-a^\dag_{\kk+\qq} a_\qq a_\kk\right)
\end{equation}
where we defined 
\begin{equation}
    \beta(k_x)\equiv \sum_{\delta} J_{\delta} \sin{(\delta k_xa)}\sin{(\delta Q)}
\end{equation}
and we symmetrized by splitting the sum in two parts and renaming the indices only in the second half; the operators are symmetric under $\kk\leftrightarrow\qq$, so they are not affected.\newline

It is now necessary to apply the same Bogoliubov transformation that makes the second-order Hamiltonian diagonal, Eq.~(\ref{eq:bogoliubov}). Let us start from the first term,
\begin{equation}
    \frac{\sqrt{2S}}{2\sqrt{N}}i\sum_{\kk\qq}\left(\beta(k_x)+\beta(q_x)\right)a^\dag_\kk a_{\kk+\qq}a^\dag_\qq
\end{equation}
Writing it in terms of the $\alpha$ operators, and collecting only the terms with two creation and one annihilation operators (the ones with two annihilations and one creation will just give the Hermitian conjugate, and there is no difference in the procedure), we get:
\begin{equation}
    \begin{split}
        \frac{\sqrt{2S}}{2\sqrt{N}}i\sum_{\kk\qq}\left(\beta(k_x)+\beta(q_x)\right)&\left(\alpha^\dag_\kk\alpha^\dag_{-\kk-\qq}\alpha_{-\qq}\sinh{\theta_{\kk+\qq}}\cosh{\theta_\kk}\sinh{\theta_\qq}\right.\\
        &\left.+\alpha^\dag_\kk\alpha_{\kk+\qq}\alpha^\dag_\qq\cosh{\theta_\kk}\cosh{\theta_{\kk+\qq}}\cosh{\theta_\qq}\right.\\
        &\left.+\alpha_{-\kk}\alpha^\dag_{-\qq-\kk}\alpha^\dag_\qq\sinh{\theta_\kk}\sinh{\theta_{\kk+\qq}}\cosh{\theta_\qq}\right)
    \end{split}
\end{equation}
There is still one step to do: we want to always have the same operator, for example $\alpha^\dag_{\kk-\qq}\alpha^\dag_\qq\alpha_\kk$. To achieve this, we use the fact that $\theta_\kk=\theta_{-\kk}$ \cite{Bogolyubov1958}, and that $\beta(k_x)$ is an odd function in $k_x$. For example, in the first term we can define a new index $\kk+\pp=\qq$:
\begin{equation}
    \sum_{\kk\pp}\left(\beta(k_x)+\beta(p_x-k_x)\right)\alpha^\dag_\pp\alpha^\dag_{\kk-\pp}\alpha_{\kk}\sinh{\theta_{\kk-\pp}}\cosh{\theta_\kk}\sinh{\theta_\pp}
\end{equation}
and then rename $\pp$ as $\qq$. After collecting all the terms, the Hamiltonian is:
\begin{equation}\label{eq:H3}
    H^{(3)}= \frac{\sqrt{2S}}{2\sqrt{N}}i\sum_{\kk\qq}C\left(\kk,\qq\right)\left(\alpha^\dag_\qq \alpha^\dag_{\kk-\qq}\alpha_\kk-\alpha_\qq \alpha_{\kk-\qq}\alpha^\dag_{\kk}\right),
\end{equation}
where we defined
\begin{equation}
    \begin{split}
        C\left(\kk,\qq\right)\equiv& \left(\beta(q_x)-\beta(k_x)\right)\left(\cosh{\theta_\kk}\cosh{\theta_{\kk-\qq}}\sinh{\theta\qq}+\sinh{\theta_\kk}\sinh{\theta_{\kk-\qq}}\cosh{\theta_\qq}\right)\\
        &-\left(\beta(k_x-q_x)+\beta(q_x)\right)\left(\sinh{\theta_\kk}\sinh{\theta_{\kk-\qq}}\sinh{\theta_\qq}+\cosh{\theta_\kk}\cosh{\theta_{\kk-\qq}}\cosh{\theta_\qq}\right)\\
        &\left(\beta(k_x)-\beta(k_x-q_x)\right)\left(\sinh{\theta_{\kk-\qq}}\cosh{\theta_\kk}\cosh{\theta_\qq}+\sinh{\theta_\kk}\cosh{\theta_\qq}\sinh{\theta_{\kk-\qq}}\right).
    \end{split}
\end{equation}

Since $\beta$ depends on $\sin{Q}$ and $\sin{2 Q}$, this interaction Hamiltonian is always zero for a collinear system, as expected. Its leading terms are linear in $Q$.

\subsection{Computational details}
Computational reasons necessitate the discretization of $k$-space. The intersection of the $k$-points on the grid and the surface where momentum and energy is conserved (Fig.~1(c)) has measure 0, so we cannot satisfy the energy conservation for any scattering processes. The widely accepted solution is to replace the $\delta$-function in energy by a Gaussian with a finite-width $\sigma_E$, i.e. allow non-conservation of energy of the order of $\sigma_E$. That allows to accumulate the value of the integral contributed by the intersection with the energy conservation surface between $k$-points $k$ and $k+\Delta k$ by summing the values at the grid sites $k$ and $k+\Delta k$. The energy change between these points is $\approx \partial E_k/\partial k \Delta k$, with $v_k=\partial E_k/\partial k$ being a magnon group velocity. Hence, we choose the spectral region of interest and a characteristic $v_k$. Since high $k$ magnons have high energy and are much less occupied, the largest contributions to $\kappa$ come from the magnons $k \lesssim Q$. Estimating the velocity from the height of the local maximum between $k=0$ and $k=Q$, we have $v=E(Q/2)/(Q/2)$ and $\sigma_E=v\Delta k=E(Q/2)2\Delta Q/Q$.
\end{document}